# Perspectives of Racetrack Memory for Large-Capacity On-Chip Memory: From Device to System

Yue Zhang, *Member, IEEE*, Chao Zhang, Jiang Nan, Zhizhong Zhang, Xueying Zhang, Jacques-Olivier Klein, *Member, IEEE*, Dafine Ravelosona, Guangyu Sun, *Member, IEEE*, Weisheng Zhao, *Senior Member, IEEE*

*Abstract*— Current-induced domain wall motion (CIDWM) is regarded as a promising way towards achieving emerging high-density, high-speed and low-power non-volatile devices. Racetrack memory is an attractive spintronic memory based on this phenomenon, which can store and transfer a series of data along a magnetic nanowire. However, storage capacity issue is always one of the most serious bottlenecks hindering its application for practical systems. This paper focuses on the potential of racetrack memory towards large capacity. The investigations covering from device level to system level have been carried out. Various alternative mechanisms to improve the capacity of racetrack memory have been proposed and elucidated, e.g. magnetic field assistance, chiral DW motion and voltage-controlled flexible DW pinning. All of them can increase nanowire length, allowing enhanced feasibility of large-capacity racetrack memory. By using SPICE-compatible racetrack memory electrical model and commercial CMOS 28 nm design kit, mixed simulations are performed to validate their functionalities and analyze their performance. System-level evaluations demonstrate the impact of capacity improvement on overall system. Compared with traditional SRAM based cache, racetrack memory based cache shows its advantages in terms of execution time and energy consumption.

## I. INTRODUCTION

Owing to the continuous increase of leakage currents, it is more and more difficult to further downscale conventional complementary metal oxide semiconductor (CMOS) [1].

Manuscript submitted on 5 Oct 2015, revised on 14 Dec 2015. This work was supported in part by the National Natural Science Foundation of China (Grant No. 61504006 and 61571023), the International Collaboration Project 2015DFE12880 from the Ministry of Science and Technology in China, the Beijing Municipal of Science and Technology (Grant No.D15110300320000), the European project FP7-MAGWIR and the French national research agency through project DIPMEM.

Y. Zhang, J. Nan, Z.Z. Zhang, X.Y. Zhang, W.S. Zhao are with School of Electronic and Information Engineering and Spintronics Interdisciplinary Center, Beihang University, Beijing, 100191, China. (E-mail: weisheng.zhao@buaa.edu.cn, Tel: +86-10-82314875).
C. Zhang and G. Sun are with Center for Energy-Efficient Computing and Applications, Peking University, Beijing, 100871, China.
J.-O. Klein and D. Ravelosona are with IEF, Univ. Paris-Sud and UMR8622, CNRS, Orsay 91405, France

Color versions of one or more of the figures in this paper are available online at http://ieeexplore.ieee.org.
Digital Object Identifier

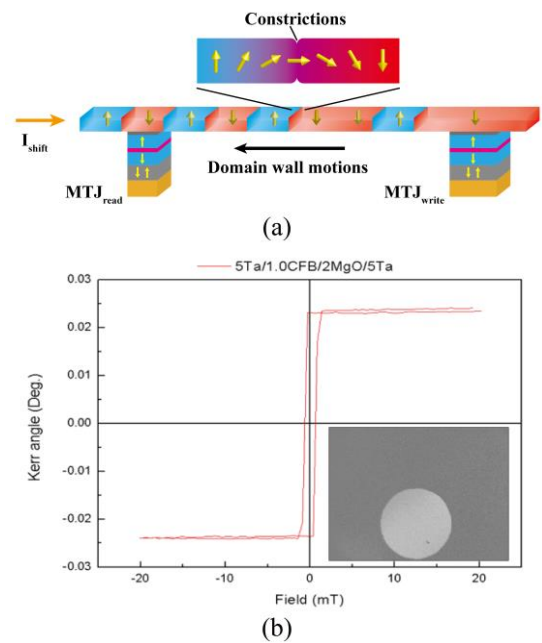

Fig. 1. (a) Racetrack memory based on CIDWM, which is composed of one write head (MTJwrite), one read head (MTJread) and one magnetic nanowire. Different data bits are separated by constrictions or notches. (b) PMA hysteresis loop of a crystallized Ta (5 nm)/CoFeB (1 nm)/MgO structure. Inset: Kerr image of perfect round magnetic domains in a crystallized Ta (5 nm)/CoFeB (1 nm)/MgO structure.

Currently CMOS based memories (e.g. SRAM and DRAM) and von Neumann architecture meet their impasse for realizing future ultra-low power computing and storage. Spintronics blazes a trail so that we can benefit from non-volatility to eliminate static energy. Recent progress of techniques and materials makes magnetic domain wall (DW) motions be regarded as a promising way to achieve non-volatile logic and memory devices [2]. In particular, due to the prospects of large density, high speed and low power, current-induced domain wall motion (CIDWM) draws further attentions from both academics and industries [3]. Compared to field-induced DW motion, CIDWM can transfer different data along one direction, which always arises from spin transfer torque (STT) effect. Thanks to this phenomenon, low power and fast computing memory application becomes expected.

Racetrack memory is one of the most attractive concepts based on CIDWM (see Fig. 1a) [4-5]. In the basic structure, a series of magnetic domains with different magnetization directions move along magnetic nanowires, driven by a spin-polarized current. Thanks to well-defined nanowires and 3D structure integration, racetrack memory demonstrates obvious potential in term of capacity. However, the path of its realization is arduous. For example, the prototype firstly fabricated was based on the materials with in-plane magnetic anisotropy, which cannot provide a sufficient thermal stability for further miniaturization [6]. In order to overcome this thermal stability issue and increase the data storage density, materials with perpendicular magnetic anisotropy (PMA) have been intensively studied (see Fig. 1b). It is also demonstrated that PMA can offer other performance improvements compared to in-plane magnetic anisotropy, such as lower DW nucleation critical current and higher DW shifting speed [7].

One more critical challenge involves the capacity bottleneck of racetrack memory, which means that each nanowire cannot be patterned so long as to store more data. This is mainly caused by two issues: high required current for DW shifting and high resistance of nanowire [8-9]. The former one is fundamentally related to the STT mechanism; the latter one is due to the material and structure design. For the given voltage supply, lower applied current and material resistance allow longer nanowire, meaning larger capacity. Indeed, many research efforts have been taken to explore how to reduce the applied current and material resistivity. For example, it has been exhibited that, due to Walker breakdown effect, magnetic field could trigger the CIDWMs below intrinsic current threshold [10]. Magnetic field can thus be one of alternative solutions to handle the problem of high DW shifting current. In addition, chiral DW motions have been discovered to allow a high efficiency for magnetization switching and DW shifting, compared to STT based CIDWM [11-14]. This phenomenon is derived from spin-orbit torque (SOT). Thanks to the low resistivity of material inducing SOT, chiral DW motions can overcome the aforementioned high resistivity drawback. Furthermore, the initial design of racetrack memory employs constrictions or notches to pin DW motions (see Fig. 1a) [15-16]. However, this could increase the resistance of nanowire and require higher current to shift DWs, which degrades the capacity performance. In this situation, alternative pinning mechanisms are urgently necessitated. Recently, voltage control of magnetic DW motions in materials with strong PMA was reported [17-18]. This phenomenon can be used to realize simultaneously flexible pinning sites and low resistance of racetrack memory device.

In this paper, we look at the potential of racetrack memory for achieving large capacity. Various PMA racetrack memory designs integrating alternative mechanisms, e.g. magnetic field assistance, chiral DW motions and voltage-controlled flexible pinning, are proposed and studied by performing mixed simulations and performance analyses using SPICE-compatible electrical models [19-21]. In addition to the device-level and circuit-level evaluations, system-level investigations are also performed. It demonstrates that capacity improvement benefits system in respects of performance and energy consumption. With an in situ replacement, racetrack memory based cache can outperform traditional SRAM based cache.

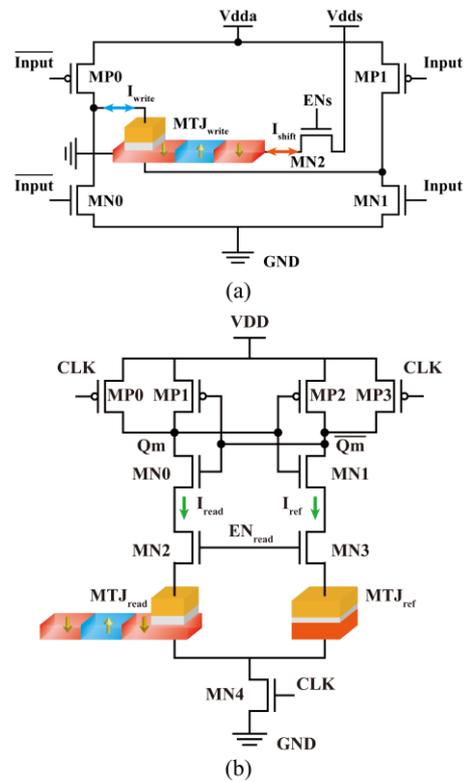

Fig. 2. (a) Writing and shifting circuits for racetrack memory. (b) Sensing circuit for racetrack memory based on pre-charge sense amplifier.

The rest of this paper is organized as follows. In section II, we introduce briefly the background and basic theory of racetrack memory. Various capacity-improving solutions will be presented in section III. The system-level study is then detailed in section IV. At last, concluding remarks and perspectives are discussed in section V.

## II. RACETRACK MEMORY INTRODUCTION

The observation of CIDWM in magnetic nanowires promises numerous perspectives and the most attractive one is to build ultra-dense non-volatile storage device, called "Racetrack memory". This term was firstly proposed by IBM in 2008 [4-5]. In this initial concept, a basic cell of racetrack memory is composed of three parts, i.e. write head, read head and storage nanowire. Write head nucleates a local domain in the magnetic nanowire. A current pulse pushes the DWs sequentially from write head to read head. The write and read heads can be constructed by magnetic tunnel junctions (i.e. MTJwrite and MTJread in Fig. 1 and Fig. 2), which facilitates hybrid integration with CMOS writing and reading circuits [22]. Data, in the form of magnetic domains, are stored along the magnetic nanowire. Different magnetic domains are separated by the transition zones, so-called DWs, which can be pinned by artificial potentials or constrictions when no current is applied. As the distance between adjacent DWs can be extremely small, this concept is possible to achieve a considerably high storage density. The nanowire can be constructed in 3D or 2D, the latter one (see Fig. 1a) is easier to be fabricated and become the mainstream solution for the current research on this topic.

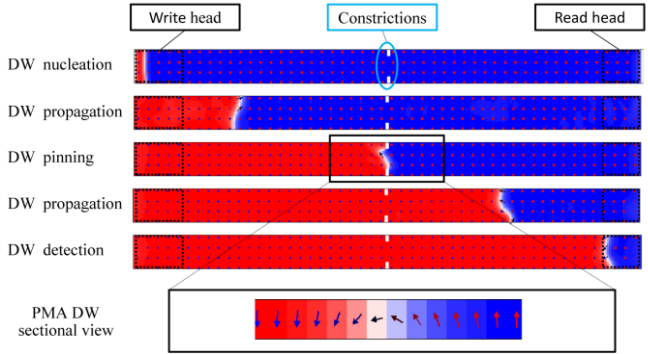

Fig. 3. Micromagnetic simulations of magnetic DW motions in a nanowire with PMA: DW is nucleated by the write head; DW can be pinned by the notch (constriction); magnetic domain is detected by the read head. The figures are vertical views of nanowire; inset shows the sectional view of PMA DW.

TABLE I. PARAMETERS AND VARIABLES INTEGRATED IN RACETRACK MEMORY WITH FIELD ASSISTANCE (CO/NI)

| Parameter | Description | Default Value |
|---|---|---|
| $\alpha$ | Gilbert damping constant | 0.045 |
| $\beta$ | Nonadiabatic coefficient | 0.02 |
| $P$ | Spin polarization rate | 0.49 |
| $M_s$ | Saturation magnetization | 0.66 MA/m |
| $H_w$ | Walker breakdown field | 4.4 mT |
| $\gamma$ | Gyromagnetic ratio | 0.176 THz/T |
| $\lambda$ | DW width | 10 nm |
| $K_u$ | Uniaxial anisotropy | 0.41 MJ/m$^3$ |
| TMR | TMR of write head MTJ | 120% |
| $J_{c\_nucleation}$ | DW Nucleation critical current density | 57 GA/m$^2$ |

With respect to structural parts, there are three indispensable currents accordingly: $I_{write}$ (see Fig. 2a) and $I_{read}$ (see Fig. 2b) execute DW nucleation and detection respectively; $I_{shift}$ (see Fig. 2a) drives DWs or data in two opposite directions. Fig. 2 also shows CMOS circuits to generate these three currents. In particular, writing circuit consisting of two PMOS and two NMOS can generate bi-directional $I_{write}$ to switch opposite states of write head. Shifting circuit can be implemented in different ways. For example, as shown in Fig. 2a, one transistor is used, and by applying positive or negative voltages on Vdds, a bi-directional current can be realized. This circuit can also be designed with two transistors connected respectively with two ends of racetrack memory. In this case, the negative voltage is not necessary and the shift direction depends on the varying voltages applied on these two ends. $I_{read}$ is driven by a sense amplifier that can convert the stored data from different magnetization orientations to digital '0' and '1'. In order to obtain the best write and read reliability, the width of write and read heads are normally different. For writing, a lower resistance of MTJwrite with larger width can reduce the risk of oxide barrier breakdown, which is one of the most significant constraints of the fast STT switching. In contrast, as the reading current is always smaller, high resistance of the MTJread with smaller width can greatly improve the sensing performance.

Thanks to its multi-bit storage ability and scalability potential, considerable efforts from academics and industries have been put into the investigation of racetrack memory. The first prototype has been successively fabricated despite of its limited capacity of 32x8 bits [6]. Meanwhile, this prototype was based on NiFe material with in-plane magnetic anisotropy, which cannot provide a sufficient thermal stability for long data retention in advanced technology nodes below 40 nm. This drawback limits its application for high-density target.

Aiming to tackle the thermal stability issue caused by in-plane anisotropy, materials with PMA were observed to offer higher energy barrier and other advantages in terms of critical current, switching speed and power consumption. With this background, racetrack memory based on magnetic nanowire with PMA (e.g. CoFeB/MgO, Co/Ni, etc.) was proposed [20]. Thanks to PMA, the distance between adjacent DWs can be decreased below 90 nm, which could further upgrade the storage density. In order to confirm the functionality of racetrack memory and optimize it with other emerging effects, we develop micromagnetic (see Fig. 3) and SPICE-compatible electrical models [19-21]. Several critical physics are integrated in them, for example, one-dimension (1D) model to describe DW motions shown as follows:

$$\dot{\varphi}_0 + \alpha \dot{X}/\lambda = \gamma H + \beta u/\lambda + f_{pin} \quad (1)$$

$$\dot{X} - \alpha\lambda\dot{\varphi}_0 = v_\perp \sin 2\varphi_0 + u \quad (2)$$

where $X$ is the position of a DW, and $\varphi_0$ is the angle that the DW magnetization forms with the easy plane. $\lambda$ is the width of DW, $\alpha$ is the Gilbert damping constant, $\beta$ is the dissipative correction to the STT, $H$ is the external field, $\gamma$ is the gyromagnetic ratio, $f_{pin}$ is the pinning force. The velocity constant $v_\perp$ comes from the hard-axis magnetic anisotropy $K_u$. $u$ is spin current velocity. Moreover, numbers of experimental parameters (shown in Tab. I and II) have also been involved. By using the electrical models, ones can design and analyze more complex logic and memory circuits [23-25].

### III. ALTERNATIVE MECHANISMS FOR CAPACITY IMPROVEMENT OF RACETRACK MEMORY

Besides the thermal stability issue, the capacity limitation of each nanowire is the most fatal bottleneck hindering practical application of racetrack memory, e.g., on-chip memory. Indeed, high current density required for CIDWMs and high resistance of magnetic nanowire are two key challenges. In the following sub-sections, we will present various alternative methods to improve capacity.

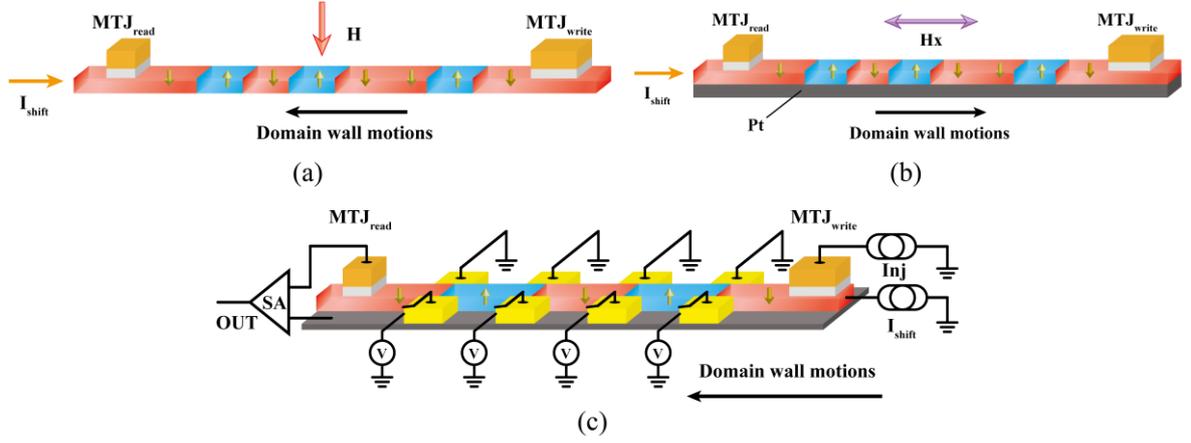

Fig. 4. Alternative mechanisms for capacity improvement of racetrack memory. (a) Magnetic field assisted racetrack memory. (b) Chiral DW motions based racetrack memory. (c) Voltage-controlled racetrack memory.

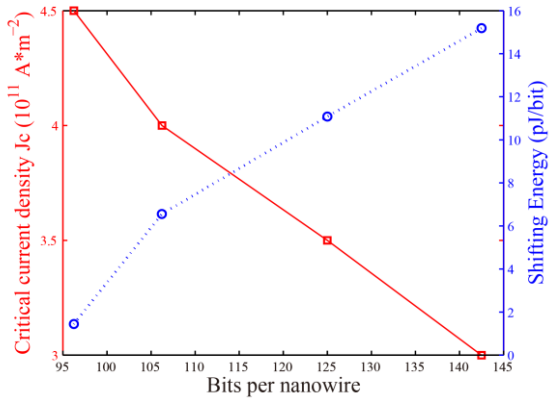

Fig. 5. Dependence of critical current density required and shifting energy versus different number of bits per nanowire in magnetic field assisted racetrack memory.

## A. Magnetic Field Assistance

Recent experimental progress showed that magnetic field would trigger DW motions below intrinsic current threshold due to Walker breakdown effect [10]. In the Co/Ni nanowire structure, the domain wall depinning should overcome the energy barrier between Neel and Bloch walls, which also determines the current threshold. Walker breakdown triggered by magnetic field could help to overcome this intrinsic energy barrier and then reduce the required current density for DW motions. This observation offers a promising orientation, i.e. magnetic field assisted racetrack memory, to relieve the pressure from high current density (see Fig. 4a). Through adding metal lines or coils, an approximately global magnetic field is generated, i.e. magnetic field H shown in Fig. 4a.

By considering the impact of magnetic field, the DW velocity is the vector sum of field-induced and current-induced velocities,

$$V = V_H + V_j \quad (3)$$

$$V_H = \alpha^2 \mu H \left\{ 1 - \frac{1}{1+\alpha^2} \sqrt{1 - \left(\frac{H_W}{H}\right)^2} \right\} \quad (4)$$

$$V_j = \frac{\mu_B P j_{sh}}{e M_S} \quad (5)$$

where the mobility $\mu = \gamma\lambda/\alpha$, $\gamma$ is the gyromagnetic ratio, $H_w$ is the Walker breakdown field. $\mu_B$ is the Bohr magneton, $e$ is the elementary charge, $P$ is the spin polarization percentage of the tunnel current and $M_S$ is the demagnetization field. It is noteworthy that the value of DW width (10 nm) is consistent with the theoretical calculation $\lambda = \sqrt{A/K_{eff}}$, where $K_{eff} = K_u - \mu_0 M_S^2/2$, $K_u$ is the uniaxial anisotropy, $\mu_0$ is the permeability in free space.

Due to this field, the current threshold for DW motions in Co/Ni material could be reduced to $3.2 \times 10^{11}$ A/m$^2$ rather than the intrinsic one, $4.5 \times 10^{11}$ A/m$^2$. By fixing the voltage supply to 3 V and the distance between two adjacent DWs to 40 nm, the limit value of bits per nanowire can exceed 100 and reach up to 150. However, the generation of magnetic field deteriorates overall power consumption. Based on the parameters shown in Tab. I, Fig. 5 shows our analyses about the relation among storage capacity, critical shifting current and power consumption. We can find that there is a tradeoff when improving the storage capacity: more bits stored per racetrack memory require lower critical current and more energy consumption. Considering different application requirements, a dual functional memory design can be implemented: large-capacity applications employ the magnetic field assistance and low-power application eliminates this assistance.

## B. Chiral DW Motion

SOT based chiral DW motions present an unprecedentedly efficient magnetization switching and high shifting speed. Opposite to the STT based CIDWMs, the direction of chiral

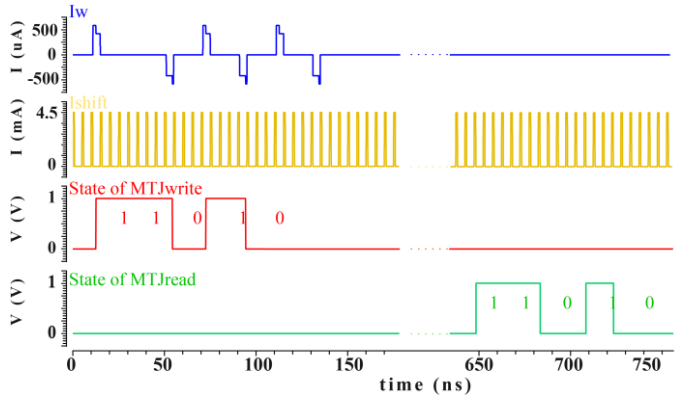

Fig. 6. Transient simulation of 64-bit racetrack memory based on chiral DW motions. Iw is used for inputting data, Ishift is used for DW shifting.

TABLE II. PARAMETERS AND VARIABLES INTEGRATED IN CHIRAL RACETRACK MEMORY (PT/CO/NI/CO)

| Parameter | Description | Default Value |
|---|---|---|
| $\alpha$ | Gilbert damping constant | 0.1 |
| $P$ | Spin polarization rate | 0.49 |
| $M_s$ | Saturation magnetization | 0.66 MA/m |
| $H_k$ | Perpendicular anisotropy field | 120 mT |
| $t$ | Ferromagnetic layer thickness | 1.15 nm |
| $\Delta$ | DW width | 3 nm |
| $D$ | DMI constant | 0.5 |
| $\theta_{SH}$ | Spin hall angle | 0.1 |

DW motions is the same with that of applied current. Although it still needs a relatively high current density for DW motions, the nanowire length can be increased thanks to the low resistivity of current-flowing non-magnetic materials (e.g., Pt, Pd, etc.) [26-27]. Therefore, chiral DW motions could be another promising solution to improve the capacity of racetrack memory. By adding a non-magnetic layer (usually heavy metal) underneath the magnetic storage nanowire, the concept of racetrack memory based on chiral DW motions is illustrated in Fig. 4b.

With regards to the impact from spin Hall effect (SHE), Dzyaloshinskii-Moriya interaction (DMI) and a longitudinal field Hx, the 1D model can be extended as follows,

$$\alpha \dot{X} + \Delta \dot{\varphi} = -\beta u - \frac{\pi}{2}\gamma \Delta H_{SHE}\sin(\varphi) \quad (6)$$

$$\dot{X} + \alpha \Delta \dot{\varphi} = \frac{\gamma \Delta H_K}{2}\sin(2\varphi) - u + \frac{\pi}{2}\gamma \Delta (H_X + H_{DMI})\sin(\varphi) \quad (7)$$

where $H_{SHE}=u\theta_{SHE}/\gamma Pt$ is the effective field describing the SHE, $t$ is the thickness of ferromagnetic layer, $\theta_{SHE}$ is the spin hall angle. $H_K$ is the anisotropy field. $H_{DMI}=D/\mu_0 M_S \Delta$ is the effective field describing the DMI, $D$ is the DMI parameter.

In the steady state, the DW velocity $v_{DW}=\dot{X}$ is analytically expressed by

$$v_{DW} = \pm \frac{\gamma \pi \Delta}{2}(H_x + H_{DMI})\frac{\frac{\pi}{2}H_{SHE}}{\frac{\pi}{2}H_{SHE} + \alpha H_K} - u\frac{\frac{\pi}{2}H_{SHE} \pm \beta H_K}{\frac{\pi}{2}H_{SHE} \pm \alpha H_K} \quad (8)$$

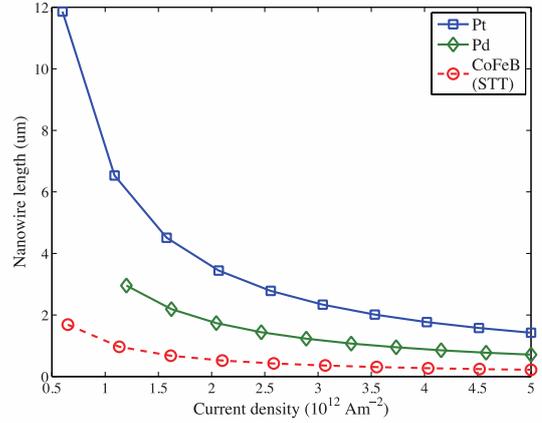

Fig. 7. Dependence of nanowire length versus DW shifting current density for Pt and Pd. Dash line shows the case for the conventional STT based CoFeB racetrack memory.

where the signs correspond to different DW configurations. According to previous experimental measurements, the DW velocity can reach as high as 300 m/s, which can be reproduced correctly by the physical models explained above.

Extraordinarily, different types of DWs (up/down or down/up) have different velocities. In certain specific conditions, one type of DWs can be halted while the other type can be accelerated. This behavior is called "peristaltic moves". Its functionality is validated with transient simulation of a 64-bit example. As shown in Fig. 6, Iw is the current applied at the write head (MTJwrite), by which a series of data "…11010…" have been input. Here, anti-parallel state represents digital '1' and parallel state represents digital '0'. After 64 pulse of Ishift, the data inputted from the write head can totally be detected by read head. In order to reveal its prospect for capacity improvement, we compare two non-magnetic materials, Pt and Pd, with conventional ferromagnetic material, CoFeB. From the result shown in Fig. 7, we can conclude that the proposed racetrack memory design can provide longer nanowire than conventional one with the same current density. Especially, Pt, with the smallest resistivity (1.56 x $10^{-7}$ $\Omega$.m), exhibits obviously the best performance among all. Considering 50 nm as the distance between adjacent DWs, chiral DW motion based racetrack memory with Pt can achieve 256 bits/nanowire.

*C. Flexible DW Pinning mechanism*

DW pinning is significant for DW manipulation and practical application. DW pinning in the initial concept of racetrack memory is carried out by constrictions or notches along the magnetic nanowire (see Fig. 1a). As these permanent defects can reduce the sectional area, the resistance of device will be dramatically increased, which is unfavorable for achieving large capacity. Flexible DW pinning mechanisms become a breakthrough for this issue.

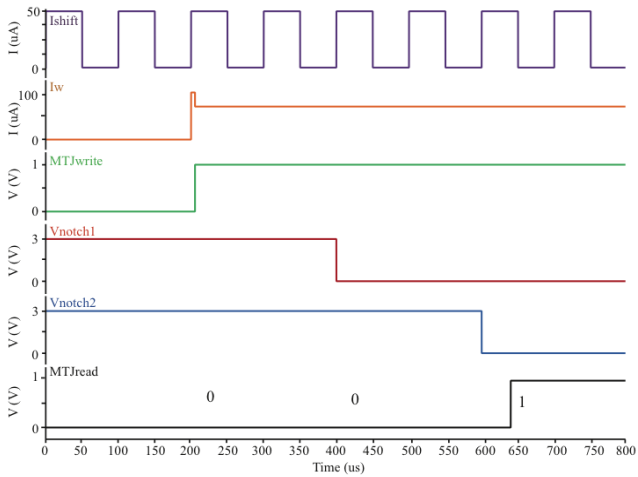

Fig. 8. Voltage controlled DW pinning in racetrack memory. (2 pinning sites).

Recently, voltage or electric field control of magnetic DW motions in materials with strong PMA was reported [17-18]. Although the control behavior has not been fully understood, the principle is always elucidated by tuning the DW motion velocity or even halting DW motions via modulation of magnetic anisotropy. As it needs gate voltage instead of current, this mechanism allows enhanced power saving and flexibility. As shown in Fig. 4c, a voltage controlled racetrack memory is proposed. The DWs are pinned by applying voltage on the gates.

The dependence of voltage on magnetic anisotropy can be assumed by

$$H_K(V) = H_{K0}(1+LV) \quad (9)$$

where $L$ is a coefficient showing the strength of the voltage-controlled anisotropy change, $H_{K0}$ is the initial anisotropy field and $V$ is the applied voltage. It should be mentioned except for some specially patterned structures, the obvious voltage effects were so far observed in creep regime. Thereby, the current-induced DW motions in creep regime are considered in our proposal. In this regime, dynamics of DW motions follows thermally activated Arrhenius law. The DW motion velocity is thus expressed as

$$v = v_0 \exp(-E_A / k_B T(1 - I/I_C)) \quad (10)$$

where $E_A = M_S H_K Vol$ is energy barrier, $v_0$ is the attempt velocity, $I_c$ is critical current. By integrating Eq.9 into Eq. 10, the impact of voltage can be taken into account.

By integrating these physics into our SPICE-compatible model of racetrack memory, the DW pinning behavior is confirmed through transient simulation (see Fig. 8). Once all the pinning gate voltage are disable, the magnetic domain can be transferred from write head to read head. However, the major challenge of this device is its operational speed. As shown in the simulation results, if the DW motion velocity is about $10^{-3}$ m/s (in creep regime), the frequency of this logic gate cannot be higher than 10 MHz (at 40 nm technology node). Recent experimental progress show that some special materials and structures can generate unprecedentedly strong voltage effects [28-29], for example, halting a DW motions with speed of 20 m/s, in flowing regime.

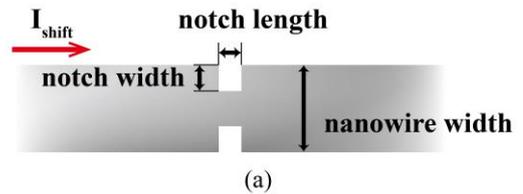

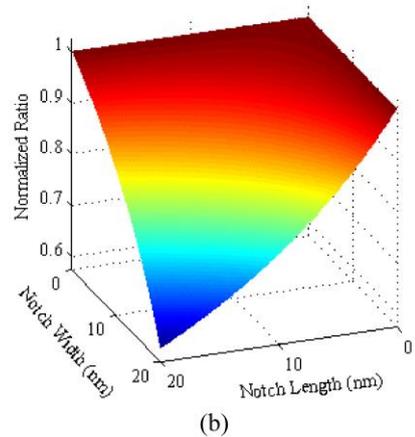

Fig. 9. (a) Rectangular notch on nanowire. (b) Dependence of size of notches on the capacity of racetrack memory

TABLE III. PERFORMANCE AND PROSPECT OF ALTERNATIVE CAPACITY-IMPROVING MECHANISMS

|  | Speed | Power | Capacity | Integration |
|---|---|---|---|---|
| Magnetic Field Assisted [10] | High (~10 m/s) | High (~10 pJ/bit) | Large (~128 bits) | Easy |
| Chiral DW Motion [14] | Ultra-High (~200 m/s) | Low (~0.1 pJ/bit) | Ultra-Large (~256 bits) | Easy |
| Voltage Controlled [28] | High (~20 m/s) | Medium (~1 pJ/bit) | Medium (~64 bits) | Medium |

In addition, we study the impact of elimination of pinning notches on capacity improvement from the resistance point of view. Considering the notches with the rectangle form (see Fig. 9a), the size of notch (width and length) will influence the resistance of nanowire. If the voltage supply is fixed, elimination or miniaturization of notches allow a longer nanowire, i.e. larger capacity. From Fig. 9b, when the width of the nanowire and the distance of notches are both fixed at 20 nm, elimination of notches can provide ~1x capacity improvement. This also demonstrates the advantage of flexible pinning mechanism on the capacity.

*D. Disscussion*

Thanks to its outstanding speed and density advantages, racetrack memory is regarded as a promising candidate for on-chip memory. With regard to the capacity challenges impeding its applications, a variety of mechanisms have been presented and investigated. However, certain additional elements (e.g.

magnetic field) have been applied, which will inevitably lead to difficulty of on-chip integration and more power consumption. Tab. III summarizes the performances and prospects of these mechanisms. From the table, as its additional element is easier to be implemented in comparison with other mechanisms, chiral DW motion based mechanism reveals the most potential to realize on-chip memory.

## IV. SYSTEM LEVEL EVALUATION

To demonstrate the benefits of using racetrack memory and improving its capacity at the system level, we evaluate the system performance based on racetrack memory cache. As the first step, we evaluate the circuit-level design of racetrack memory. At the system level, we compare the time and energy performance of racetrack memory with different capacities. According to the device-level optimization mechanisms proposed in the previous section, the capacity limit of each case has been considered. At last, comparison with conventional SRAM based cache is explored and discussed.

### A. Racetrack memory circuits

In order to exploit racetrack memory as on-chip memory, we need first implement a random access memory (RAM) in synchronous circuit. The circuit design follows previous work [30], based on NVsim [31]. We build racetrack memory based RAM to fit cache design. The outline of a memory bank is shown in Fig. 10. Each bank includes several mats, and each mat has several arrays. The basic storage unit of an array is called cell, which is the nanowire of racetrack memory described in previous sections. Besides array of cells, there are row decoders, domain decoders, pre-chargers, word line drivers, column multiplexers, sense amplifiers, output drivers and extra logic as periphery circuitry to operate the storage.

In order to improve the capacity of racetrack memory, several alternative mechanisms have been proposed. However, supplying extra electrical or magnetic field in memory circuit causes extra design effort and overhead. Even though the performance of a RAM depends largely on the characteristics of the cell, the periphery circuitry always still induces a large portion of overhead (e.g. ~20% of area, ~40% of latency). To apply magnetic field assistance method, coupling lines under the layer of racetrack nanowire should be supplied with current to provide required magnetic field. To apply voltage-controlled method, pair of electrons routed by signal lines should be introduced. To implement chiral DW motion method, extra conducting lines should be bound with the magnetic nanowire, which needs more metal layers. To make these methods reliable, the energy and side effect of these lines should be seriously modeled. A more accurate simulation of the circuit should also take these extra design overheads into account, which is out of the scope of this work.

### B. Cell Capacity Improvement

Capacity improvement is the focus of this paper, the influence of racetrack memory RAM capacity at system level should be investigated. Since abovementioned mechanisms can increase the storage capacity of each nanowire, we first simulate the system performance and energy consumption after increasing the cell capacity. We use gem5 [32], a cycle accurate full system simulator, to simulate PARSEC [33] benchmark suite for performance and power evaluation. All benchmarks are fully executed and we count the overall execution time of the cache system to evaluate the performance. We consider both static power and dynamic energy to evaluate energy consumption. We collect the operation amount and execution time of each benchmark from gem5 simulator, and calculate the energy with parameters from previous work [30]. The system is configured with 4 simple cores, private 32KB I/D cache, and a shared L2 cache. The configurations are shown in Tab. IV.

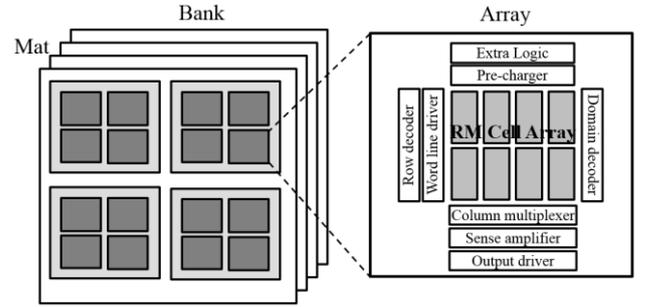

Fig. 10. The design outline of a racetrack memory RAM.

TABLE IV. SYSTEM LEVEL EXPERIMENT SETUP DETAILS

| Component | Configuration |
|---|---|
| Processor | 4 simple cores, 2GHz, 1-way issue |
| L1 I/D Cache | 32/32KB, 2-way, 64B line, private, LRU SRAM, 1/1 cycle, 6.2/2.3pJ |
| L2 Cache | 16-way, 64B, shared, LRU SRAM: 2MB, 10/10 cycle, 0.57/0.54nJ, 3438mW RM: 64-192MB, 9/20/5 cycle, 0.50/0.55/0.50nJ, 1062mW |
| Main Memory | 8GB, DDR3, 1600MHz, 120cycle, 12.8GB/s |

Different numbers of bit per racetrack memory have been involved in Fig. 11: 32 bits/nanowire and 64 bits/nanowire can be achieved by conventional STT based racetrack memory and flexible pinning mechanism; 128 bits/nanowire can be reached by the magnetic field assistance mechanism; 256 bits/nanowire can be obtained through chiral DW motion mechanism. When the capacity of a cell is increased to 128 from 32, the performance upgrades by 5% on average. But the performance gain reduces when the capacity keeps increasing. This is because increasing the capacity also increases the shift latency, and hence degrades the performance. However, thanks to the high speed of chiral DW motion, i.e. as high as 300 m/s, the shift latency of this mechanism can greatly be shortened. As a result, the chiral DW motion based racetrack memory shows the shortest overall execution time and retain the performance gain.

Then we study the effect of cell capacity on energy consumption. As mentioned in the Section. III, 256 bits/nanowire can be achieved. We analyze this effect from 32 bits/nanowire to 256 bits/nanowire, as shown in Fig. 12. Because the shift energy per bit is reduced, larger capacity indeed saves more energy. It is noteworthy that, if the additional forces, e.g. magnetic field, are taken into account, the overall energy consumption will be degraded. Nevertheless, this study still exhibits the benefit of capacity improvement for energy saving.

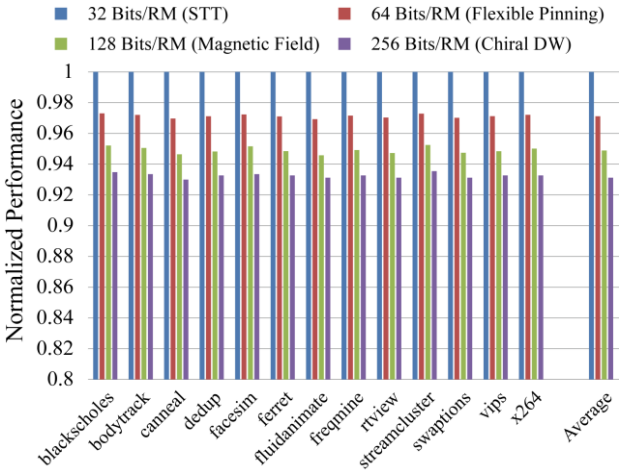

Fig. 11. Comparison of system overall execution time among variant cell capacity. (RM: racetrack memory)

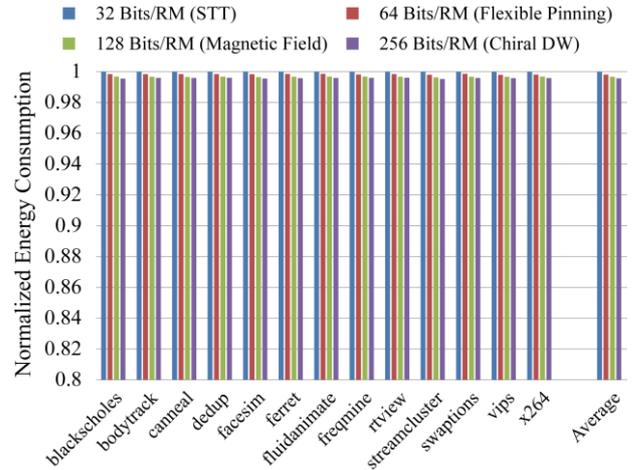

Fig. 12. Comparison of L2 cache energy among variant cell capacity. (RM: racetrack memory)

Besides the system performance and energy, increasing the capacity of racetrack memory cell also enlarges the circuit-level design space. One method proposed to tolerate position errors in shift operations is the position error correction codes (p-ECC) [34]. This code significantly relies on the capacity of the nanowire. Its error check overhead on area is 17.6%. And it could be reduced to 9%, if the capacity is doubled. Additionally, racetrack memory is also employed as a part in-memory AES encryption engine [35], which utilizes the shift operations to perform matric row shift calculation. The increase of capacity will definitely expand the matric size that can be calculated and improves the calculation speed.

### C. Memory Capacity Improvement

We then compare racetrack memory based cache with different capacities in terms of performance and energy consumption. Fig. 13 outlines the overall execution time change with increased L2 cache capacity. Since traditional programs are optimized for small on-chip cache (e.g., 4MB), they can hardly benefit from ultra-large on-chip cache (e.g., 192MB). Thus the most benchmarks show trivial performance improvement. In addition, the overall effect is partially covered by instruction level parallelism. Even though the latency of cache access becomes larger for larger caches, the benchmarks that have relative lower cache access intensity demonstrate almost the same performance. But memory intensive applications (e.g. Facesim, cannel) still get significant performance improvement. The execution time reduces by 3% and 8% for facesim, after using 128MB and 192MB L2 cache. The average overall execution time reduction for using 128MB and 192MB L2 cache is 1% and 2%. The overall cache energy shows similar result, shown in Fig. 14. For applications requiring large memory, the energy consumption of cache is reduced by employing larger cache. The average overall cache energy is reduced by 2.3% and 3.1% after using 128MB and 192MB L2 cache compared to 64MB L2 cache.

### D. Comparison with SRAM

To fully demonstrate the benefits of using racetrack memory at the system level, we also compare conventional SRAM based L2 cache with racetrack memory based cache in Fig. 13 and Fig. 14. In order to compare an on situ replacement, we conduct an iso-area comparison. For example, 2MB SRAM costs ~6.5 mm$^2$ area, which is nearly equivalent to 64MB conventional STT based racetrack memory area (~6.2 mm$^2$).

The overall system performance is shown in Fig. 13. Due to the increase of cache capacity, some benchmarks (e.g. canneal) show significant performance improvement. But some benchmarks (e.g. ferret) suffer from performance degradation due to extra shift latency, which induces about 7.5% overhead. The maximum performance improvement can be up to 38.3%, and the average improvement is 15.8%.

The results of energy consumption are shown in Fig. 14. Generally, racetrack memory reduces the energy consumed in L2 cache. Even though the dynamic energy for racetrack memory is higher than SRAM, the static leakage power of racetrack memory is much lower. The replacement will finally save the system power consumption. It reduces the L2 cache energy up to 79.3%, and 73.0% on average.

In summary, increasing the capacity benefits the system-level design, and indeed expands the design space for racetrack memory. Among the proposed mechanisms, chiral DW motion based racetrack memory shows the best prospect

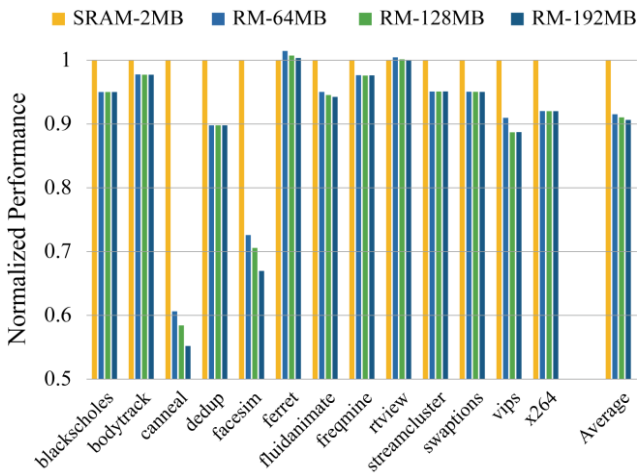

Fig. 13. Comparison of system overall execution time among SRAM and racetrack memory with different capacities. (RM: racetrack memory)

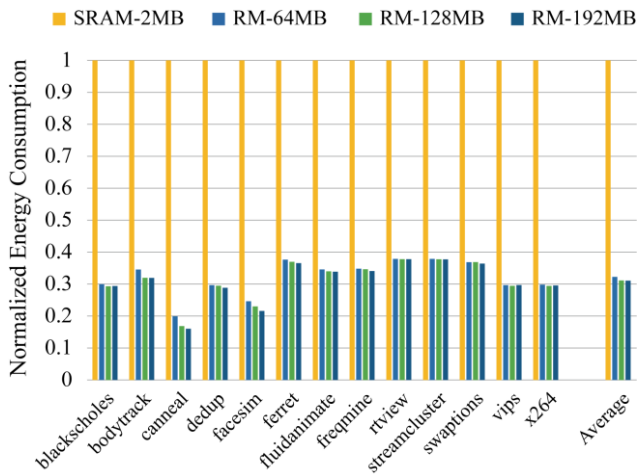

Fig. 14. Comparison of L2 cache energy among SRAM and racetrack memory with different capacities. (RM: racetrack memory)

from the points of view of device-level capacity improvement and system-level performance.

## V. CONCLUSION

This paper presents the prospect of racetrack memory through the viewpoints from device design to system evaluation. Firstly, alternative ways to improve capacity have been proposed and summarized. Magnetic field assistance taking advantage of Walker breakdown effect could reduce required current density. Chiral DW motion provides a new mechanism to achieve racetrack memory and the corresponding materials possess lower resistivity than those for conventional STT based CIDWMs. Voltage-controlled magnetic anisotropy effect enables resistance reduction and length optimization by avoiding the employment of permanent pinning sites. The improvement of capacity could make racetrack memory device more appropriate for those systems that require large on-chip memory. From system-level evaluations, racetrack memory also shows its advantages over traditional memories in terms of program execution time and energy consumption.

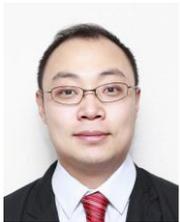

**Yue Zhang** (S'11, M'14) received B.S. in optoelectronics from Huazhong University of Science and Technology, Wuhan, China, in 2009, M.S. and Ph.D. in microelectronics from University of Paris-Sud, France, in 2011 and 2014, respectively. He is currently an associate professor at Beihang University, China.

His current research focuses on emerging non-volatile memory technologies and hybrid low-power circuit designs. He has authored more than 50 scientific papers in referred journals and conferences and received several best paper awards from the international conferences, such as NANOARCH, NEWCAS and ESREF.

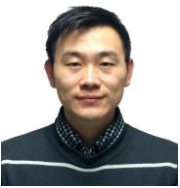

**Chao Zhang** received his B.S. degree from Peking University, China, in 2012. He is pursuing Ph.D. degree in Peking University, advised by Dr. Guangyu Sun. His research interests include computer architecture, racetrack memory design. He is a student member of IEEE, CCF.

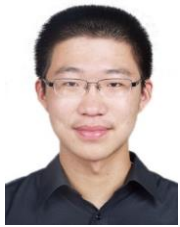

**Jiang Nan** was born in China in 1990. He received his B.S. degree in electronic and information engineering from Beihang University, Beijing, China, in 2008. He is working for his Ph.D degree at Beihang university. His current research interests include simulation analysis of MTJ and all spin logic device.

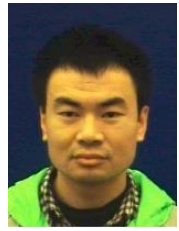

**Zhizhong Zhang** was born in China, in 1990. He received the B.S. degree from Beihang University. He is currently working toward the Ph.D. degree in microelec- tronics at Beihang University, Beijing, China. His current research interests include the theoretical magnetism and micromagnetic simulation.

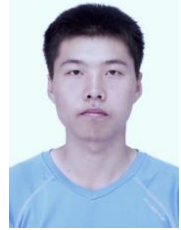

**Xueying Zhang** was born in China, in 1987. He received the B.S. and M.E. degrees from the Ecole Centrale de Pekin of Beihang University, Beijing, China, respectively in 2011 and 2014. He is currently working toward the Ph.D. degree in microelectronics at Beihang University, Beijing, China.His current research interests include the domain wall motion in ferromagnetic nanowire, the excitation and propagation of spin wave, magneto dynamic measurements via magneto-optical Kerr effect.

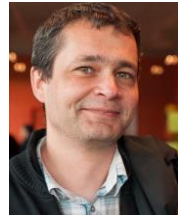

**Jacques-Olivier Klein** (M'90) was born in France in 1967. He received the Ph.D. degree and the Habilitation in electronic engineering from the University Paris-Sud, France, in 1995 and 2009 respectively.He is currently professor at University Paris-Sud, where he leads the nano-computing research group focusing on the architecture of circuits and systems based on emerging nanodevices in the field of nano-magnetism and bio-inspired nano-electronics.

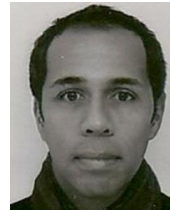

**Dafiné RAVELOSONA** is an experimentalist physicist and he is currently the head of the"Nanospintronics" group at Institut d'Electronique Fondamentale (IEF), in Orsay (France). After a Ph. D. in Solid States Physics (1995) and a post doc fellowship at CNM in Madrid (Spain), he became a permanent research member of CNRS in 1998 at the University of Paris Sud. From 2004 to 2005, he joined as an invited scientist the research center of Hitachi Global Storage Technology, San José, USA. His work has mainly focused on transport phenomena in nanostructures with perpendicular anisotropy for applications to logic and solid state memories.

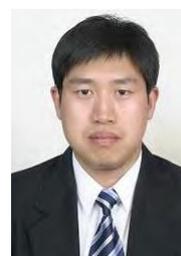

**Guangyu Sun** received the B.S. and M.S. degrees in Electronic Engineering and Microelectronics from Tsinghua University, Beijing, in 2003 and 2006, respectively. He received his Ph.D. degree in Computer Science from Pennsylvania State University, State College, PA in 2011. He is currently an assistant professor in the Center for Energy-efficient Computing and Applications, Peking University. His research interests include computer architecture, storage system, and electronic design automation. He is a member of CCF, IEEE, and ACM.

**Weisheng Zhao** (M'06, SM'14) received the Ph.D. degree in physics from the University of Paris-Sud, France, in 2007.

From 2004 to 2008, he investigated Spintronic devices based logic circuits and designed a prototype for hybrid Spintronic/CMOS (90 nm) chip in cooperation with STMicroelectronics. Since 2009, he joined the CNRS as a tenured research scientist and his interest includes the hybrid integration of nanodevices with CMOS circuit and new non-volatile memory (40 nm technology node and below) like MRAM circuit and architecture design. He has published more than 150 scientic articles in the leading journals such as Nature Communications, Advanced Materials and conferences such as IEEE-DATE, ISCA etc. Since 2014, he becomes distinguished professor in Beihang University, Beijing, China. From 2015, he becomes associate editor of IEEE Transactions on Nanotechnology, Electronics Letters and guest editor of IEEE Transactions on Multi-scale computing system.